\documentclass[aps, pra, reprint, showpacs, amsmath, amssymb]{revtex4-2}

\usepackage{amsmath,amscd}
\usepackage{bm}% bold math
\usepackage[english]{babel}
\usepackage{graphicx}% Include figure files
\usepackage{soul}
\usepackage[usenames]{color}
\begin{document}
\title{ Cherenkov Radiation from a Hollow Conical Targets: Off-Axis Charge Motion }

\author{ Andrey V. Tyukhtin }
\email{a.tyuhtin@spbu.ru}
\author{ Sergey N. Galyamin }
\affiliation{Saint Petersburg State University, 7/9 Universitetskaya nab., St. Petersburg, 199034 Russia}
\author{ Viktor V. Vorobev }
\affiliation{Technical University of Munich, Germany, Department of Informatics}

\date{\today}

\begin{abstract}
Cherenkov radiation (CR) generated by a charge moving through a hollow conical target made of dielectric material is analyzed. 
We consider two cases: the charge moves from the base of the cone to its top (``straight'' cone) or from the top to the base (``inverted'' cone). 
Unlike previous papers, a nonzero shift of the charge trajectory from the symmetry axis is taken into account which leads to generation of asymmetric CR. 
The most interesting effect is the phenomenon of ``Cherenkov spotlight'' which has been reported earlier for axially symmetric problems. 
This effect allows essential enhancement of the CR intensity in the far-field region by proper selection of the target's parameters and charge velocity. 
Here we describe the influence of charge shift on CR far-field patterns paying the main attention to the ``Cherenkov spotlight'' regime. 
Influence of variation of the charge speed on this phenomenon is also investigated. 
\end{abstract}

\maketitle

%%%%%%%%%%%%%%%%%%%
%%%%%%%%%%%%%%%%%%%
\section{Introduction\label{sec:intro}}
%%%%%%%%%%%%%%%%%%%
%%%%%%%%%%%%%%%%%%%

Cherenkov radiation (CR) discovered in 1937~\cite{Ch37} has found a series of well-known applications in high-energy physics connected mainly with particle detection~\cite{Jb, Zrb}.
In addition, in recent years several other areas of contemporary physics are tightly connected with CR effect.

For example, CR in the form of a ``wakefield'' generated by relativistic charged particle bunches in dielectric-lined waveguides is considered as prospective for both particle acceleration (within ``wakefield acceleration'' scheme~\cite{JingAntipov18}) and development of beam-driven radiation sources~\cite{GTAB14, OShea16, Ant16, WangAntipov2018}.
Cherenkov-type radiation produced by moving pulses of polarization (these pulses occur due to the optical rectification of laser pulses in nonlinear medium) is widely used for generation of Terahertz (THz) radiation~\cite{Hebling2019Review, Hebling2020}.
Prospective modification of this scheme consists in dividing a nonlinear ``core'' where a traveling polarization source propagates and a radiator (dielectric ``target'' made of linear material) where a THz radiation is generated by this source~\cite{Bakunov2012, Bakunov2014, BakunovBodrov2020}.
Similar dielectric targets of complicated form are also of essential interest for beam-driven THz radiation sources~\cite{Sei15, Sei17, Potylitsyn2020SciRep} and bunch diagnostics~\cite{Kieffer18, Kieffer2020}. 

``Opened'' dielectric radiators (not a waveguide structures) used in the aforementioned papers typically have their sizes much larger than the wavelengths under consideration. 
On the one hand, this fact complicates considerably computer simulations due to the large amount of required resources. 
On the other hand, this fact allows introducing an obvious small parameter of the problem and therefore developing corresponding approximate methods of analysis.
In a series of our preceding papers, we have presented two efficient methods based on the mentioned idea. They have been called a``ray-optics method'' and an ``aperture method'': fundamentals of these approaches can be learned from~\cite{TGV19, GV20PRAB, TGVG20_PRA, TGV21_JOSAB}.
The ``aperture method'' is more general (compared to the ``ray optics'' one) and will be used in the present paper. 
Verification of the discussed approaches via COMSOL simulations (for axisymmetric problems) has been performed in papers~\cite{GTV18, TVBG18, TBGV20}: as expected, the accuracy of the ``aperture method'' is of order of the ratio between the wavelength and object's size (at least in the regions of radiation field maxima). 
This conclusion is especially important in the context of the present paper because we are mainly interested in the region of the main maximum of the radiation pattern. 

Earlier, we considered two problems with a dielectric cone having a vacuum channel under condition that the charge trajectory coincides with the symmetry axis of the cone~\cite{TGV19, TGVG20_PRA}. 
One problem assumed the motion of the charge from the cone base to the cone top  (``straight'' cone), and the other dealt with the charge movement from the top to the base (``inverted'' cone). 
In both problems, the most interesting phenomenon is an effect named by us the ``Cherenkov spotlight'' effect. 
This effect allows essential increasing the CR intensity in the far-field (Fraunhofer) area by proper selection of the target's parameters and the charge velocity. 
The effect takes place if the refracted waves in the ray optics area are parallel to the symmetry axis. 
In this case the maximal radiation in the Fraunhofer area is observed at some small (but nonzero!) angle with respect to this axis. 
This radiation is much more intensive than the radiation in other situations. 

However, it is important for practice to establish how sensitive the ``Cherenkov spotlight'' effect is with respect to the shift of the charge trajectory from the symmetry axis.  
Therefore, in this paper we study the case when the charge moves parallel to the cone symmetry axis but along shifted trajectory. 
We consider both the ``straight'' cone and the ``inverted'' cone cases. 
These problems are essentially more complicated compared to the axisymmetric problems~\cite{TGV19, TGVG20_PRA} because of a cylindrical symmetry absence. 
We pay the main attention to analysis of radiation in the far-field zone, especially to the ``Cherenkov spotlight'' effect. 

%%%%%%%%%%%%%%%%%%%%%%%%%%%%
%%%%%%%%%%%%%%%%%%%%%%%%%%%%
\section{\label{sec:2} The case of the ``straight'' cone }
%%%%%%%%%%%%%%%%%%%%%%%%%
%%%%%%%%%%%%%%%%%%%%%%%%%
\subsection{\label{subsec:2.1}  The field on the aperture}
%%%%%%%%%%%%%%%%%%%%%%%%%
%%%%%%%%%%%%%%%%%%%%%%%%

\begin{figure}
\centering
\includegraphics{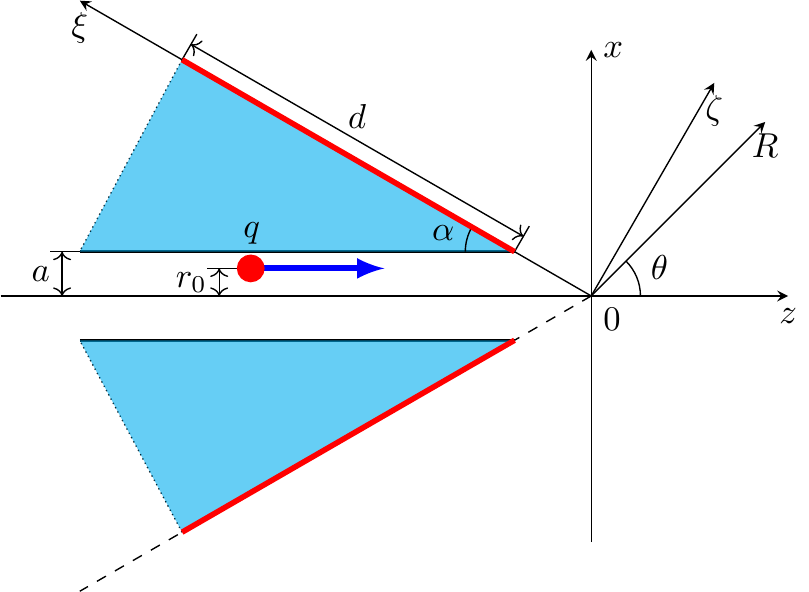}
\caption{\label{fig:1}%
The cone cross-section.
}
\end{figure}

We analyze radiation of a charge moving along the axis of the cylindrical channel with radius
$ a $ in a conical object (Fig.~\ref{fig:1}) made of dielectric material with permittivity $ \varepsilon $%
, permeability $ \mu = 1 $ and negligible conductivity. 
In this section we assume that the charge moves from the target base to its top. 
In this situation the actual form of the lateral border of the target is not principal (dotted line in Fig.~\ref{fig:1}), only the aperture with size $d$ illuminated by CR is important (red lines in Fig.~\ref{fig:1}). 
The target sizes are supposed to be much larger than wavelengths under consideration: 
$d \gg \lambda $. 
The coordinate system origin is at the cone apex, and the $ z $%
-axis is the symmetry axis of the target.
%
%%%%%%%%%%%%%%%%%%%%%%%%%%%%%%%%%%%%%%%%%%%%%%%%%%%%%%%%%%%%%%%%%%%%%%%%%%%%%%%%%%%%%%%%%
%%%%%%%%%%%%%%%%%%%%%%%%%%%%%%%%%%%%%%%%%%%%%%%%%%%%%%%%%%%%%%%%%%%%%%%%%%%%%%%%%%%%%%%%%%%%

The point charge
$ q $
moves with constant velocity
$ \vec{ \upsilon } = c \beta \vec{ e }_z $
parallel to
$ z $%
-axis along the straight trajectory shifted by
$ r_0 $
along
$ x $%
-axis. 
The charge density is 
$ \rho = q \delta ( x-r_0) \delta ( y ) \delta ( z - c \beta t ) $,
where $ \delta $ is a Dirac delta-function.
It is assumed that the charge velocity exceeds the ``Cherenkov threshold'', i.e.
$ \beta > 1 / n $,
where
$ n = \sqrt{ \varepsilon }$
is a refractive index of the target material. Therefore, CR is generated in the cone material.

We will use here the same notations as in the paper of Tyukhtin et al.~\cite{TGV19}. 
The ``main'' wave is the wave which causes the ``Cherenkov spotlight'' effect. 
The way of this wave is the same as in the axially symmetrical case $ r_0 = 0 $ (see Fig.~2 in \cite{TGV19}). 
The CR wave generated by a moving charge in the bulk of the target incidents the inner surface of the aperture at the angle 
\[
\theta_{ i } = \pi / 2 - \alpha - \theta_p,
\quad
\theta_p = \arccos \left[ 1 / \left( n \beta \right) \right],
\]
and then refracted at the angle
\[
\theta_{ t } = \arcsin\left(1/(n\beta)\right)
\]
with respect to the boundary normal $\vec{e}_\zeta$ (see Fig.~2 in \cite{TGV19}).

Due to the asymmetry caused by the charge trajectory shift, CR of both polarizations is generated. 
``Parallel'' polarization ($ \parallel $) contains components
$ E_ z$, $ E_r$ and $ H_\varphi $. 
Corresponding Fresnel transmission coefficient (for the transmission at the cone generatrix) is
\begin{equation}
\label{eq:Tpar}
T_{ \parallel}
=
\frac{ 2 \cos \theta_{ i } }{ \cos \theta_{ i } + \sqrt{ \varepsilon } \cos \theta_{ t } }.
\end{equation}
``Orthogonal'' polarization ($ \bot $) contains components
$ H_ z$,  $ H_\rho$ and $ E_\varphi$.  
Corresponding Fresnel coefficient is
\begin{equation}
\label{eq:Tort}
T_{ \bot }
=
\frac{ 2 \sqrt{ \varepsilon } \cos \theta_{ i } }{ \sqrt{ \varepsilon } \cos \theta_{ i } + \cos \theta_{ t } }.
\end{equation}

Based on results of our previous papers~%
\cite{TGV19, GV20PRAB, TGVG20_PRA}
we can write the CR field before the refraction at the cone generatrix surface. 
For $ \parallel $-polarization, the Fourier-transform of the azimuthal magnetic component at the distance $ r \gg \lambda $ is

\begin{equation}
\label{eq:Hphiappr}
\begin{aligned}
&H_{ \varphi }^{ i }
( r, \varphi,  z )
=
\frac{ - q \omega }{ \pi \upsilon^2 \gamma^2 }
\sqrt{ \frac{ 2 }{ \pi r s } }
\exp \left[ i \Phi_{ i } ( r, z ) \right]
{\times} \\
&{\times}
\frac{ i k }{ s }
\!\!
\left\{
- \varepsilon 
I_0 ( r_0 \sigma )
\tilde{ A }_0^{ ( E2 ) }
{+}
2
\sum\limits_{ \nu = 1 }^{ \infty }
I_{ \nu } ( r_0 \sigma )
e^{ \frac{ i \pi ( 1 {-} \nu ) }{ 2 } }
{\times}
\right. \\
&\left.
\vphantom{ \sum\limits_{ \nu = 1 }^{ \infty } }
\times
\cos ( \nu \varphi )
%\left[
i \varepsilon
\tilde{ A }_{ \nu }^{ ( E2 ) }
%\left( i s - \frac{ 1 }{ 2 r } \right)
%-
%\frac{ i \nu }{ \beta r }
%\tilde{ A }_{ \nu }^{ ( H2 ) }
%\right]
\right\}.
\end{aligned}
\end{equation}
For $ \bot $ polarization, the Fourier-transform of the azimuthal electric component is 
\begin{equation}
\label{eq:Ephiappr}
\begin{aligned}
&E_{ \varphi }^{ i }
( r, \varphi,  z )
=
\frac{ q \omega }{ \pi \upsilon^2 \gamma^2 }
\sqrt{ \frac{ 2 }{ \pi r s } }
\exp \left[ i \Phi_{ i } ( r, z ) \right]
{\times} \\
&{\times}
\frac{ i k }{ s }
2
\sum\limits_{ \nu = 1 }^{ \infty }
I_{ \nu } ( r_0 \sigma )
e^{ \frac{ i \pi ( 1 {-} \nu ) }{ 2 } }
%{\times} \\
%&{\times}
\sin ( \nu \varphi )
%\left[
%i
\tilde{ A }_{ \nu }^{ ( H2 ) }
%\left( i s - \frac{ 1 }{ 2 r } \right)
%-
%\frac{ \nu }{ \beta r }
%\tilde{ A }_{ \nu }^{ ( E2 ) }
%\right].
\end{aligned}
\end{equation}
Here we introduce the following notations: 
\begin{widetext}
\[
\tilde{ A }_0^{ ( E1 ) }
=
-\frac{ \sigma^2 s \varepsilon H_0^{ \prime } K_0 + \sigma s^2 K_0^{ \prime } H_0 }
{ \sigma^2 s \varepsilon H_0^{ \prime } I_0 + \sigma s^2 I_0^{ \prime } H_0 },
\]
\begin{equation}
\label{eq:AnuE1}
\tilde{ A }_{ \nu }^{ ( E1 ) }
{=}
\frac{ 1 }{ \Delta_{ \nu } H_{ \nu }^2 }
\left\{
-\left[
\nu ( \beta a )^{ -1 } I_{ \nu } ( \sigma^2 + s^2 )
\right]^2
H_{ \nu }^2 K_{ \nu } I_{ \nu }^{ -1 }
{+}
\left[
\sigma^2 s \mu H_{ \nu }^{ \prime } I_{ \nu } {+}
s^2 \sigma I_{ \nu }^{ \prime } H_{ \nu }
\right]
\left[
\sigma^2 s \varepsilon H_{ \nu }^{ \prime } K_{ \nu } {+}
s^2 \sigma K_{ \nu }^{ \prime } H_{ \nu }
\right]
\right\},
\end{equation}
\begin{equation}
\label{eq:AnuH1}
\tilde{ A }_{ \nu }^{ ( H1 ) }
{=}
\frac{ \nu I_{ \nu } ( \sigma^2 + s^2 ) }{ i \beta a \Delta_{ \nu } H_{ \nu } }
\left\{
\left[
\sigma^2 s \varepsilon H_{ \nu }^{ \prime } K_{ \nu } {+}
s^2 \sigma K_{ \nu }^{ \prime } H_{ \nu }
\right]
{-}
K_{ \nu } I_{ \nu }^{ -1 }
\left[
\sigma^2 s \varepsilon H_{ \nu }^{ \prime } I_{ \nu } {+}
s^2 \sigma I_{ \nu }^{ \prime } H_{ \nu }
\right]
\right\},
\end{equation}
\begin{equation}
\label{eq:Det}
\Delta_{ \nu }
=
\left[
\nu ( \beta a )^{ -1 }
I_{ \nu }
\left( \sigma^2 + s^2 \right)
\right]^2
-
\frac{
\left[
\sigma^2 s \varepsilon H_{ \nu }^{ \prime } I_{ \nu } {+}
s^2 \sigma I_{ \nu }^{ \prime } H_{ \nu }
\right]
\!
\left[
\sigma^2 s \mu H_{ \nu }^{ \prime } I_{ \nu } {+}
s^2 \sigma I_{ \nu }^{ \prime } H_{ \nu }
\right]
}
{ H_{ \nu }^2 },
\end{equation}
\end{widetext}
where 
$$ k = \omega / c , \ \ \ 
s = k \sqrt{ \varepsilon \beta^2 - 1 } / \beta , 
$$
$$ \sigma = k  \sqrt{ 1 - \beta^2 } / \beta=k  / \beta / \gamma , \ \ \ 
 \gamma = 1 / \sqrt{ 1 - \beta^2 }, 
$$
\begin{equation}
\label{eq:IHprimeIH}
\begin{aligned}
I_{ \nu } &\equiv I_{ \nu }( a \sigma ), &
H_{ \nu } &\equiv H_{ \nu }^{ ( 1 ) }( a s ),
\\
I_{ \nu }^{ \prime } &\equiv \left. \frac{ d I_{ \nu }( \xi ) } { d \xi }
\right|_{ \xi = a \sigma }, &
H_{ \nu }^{ \prime } &\equiv \left. \frac{ d H_{ \nu }^{ ( 1 ) }( \xi ) }{ d \xi }
\right|_{ \xi = a s },
\end{aligned}
\end{equation}
\begin{equation}
\label{eq:KprimeK}
K_{ \nu } \equiv K_{ \nu }( a \sigma ),
\quad
K_{ \nu }^{ \prime } \equiv \left. \frac{ d K_{ \nu }( \xi ) } { d \xi }
\right|_{ \xi = a \sigma }. 
\end{equation}
Note that the following recurrent formulas are useful for calculation of the derivatives:  
\begin{equation}
\label{recur}
\begin{aligned}
I_{ \nu }^{ \prime } &= I_{ \nu + 1 } + \frac{ \nu }{ a \sigma } I_{ \nu }, \\
K_{ \nu }^{ \prime } &= - K_{ \nu + 1 } + \frac{ \nu }{ a \sigma } K_{ \nu }, \\
H_{ \nu }^{ \prime } &= - H_{ \nu + 1 } + \frac{ \nu }{ a s } H_{ \nu },
\end{aligned}
\end{equation}
\begin{equation}
\label{eq:AnuE2H2}
\tilde{ A }_{ \nu }^{ ( E2 ) }
{=}
\tilde{ A }_{ \nu }^{ ( E1 ) }
\frac{ I_{ \nu } }{ H_{ \nu } }
+
\frac{ K_{ \nu } }{ H_{ \nu } },
\quad
\tilde{ A }_{ \nu }^{ ( H2 ) }
{=}
\tilde{ A }_{ \nu }^{ ( H1 ) }
\frac{ I_{ \nu } }{ H_{ \nu } }.
\end{equation}

Phase $\Phi_{ i }( r, z )$ is the phase of the CR inside the bulk of the target:
\[
 \Phi_{ i }( r, z ) = s r + \frac{ k z }{ \beta } - \frac{ \pi }{ 4 }.
\]

Further the coordinates with primes denote the coordinates at the aperture. 
Taking into account that on the generatrix surface $r^{ \prime } = \xi^{ \prime } \sin(\alpha)$ and $z^{ \prime } = - \xi^{ \prime } \cos(\alpha)$, one can get this phase the generatrix surface as
\begin{equation}
 \Phi_{ i }( \xi^{ \prime } ) = s \xi^{ \prime } \sin\alpha - \frac{ k }{ \beta } \xi^{ \prime } \cos\alpha - \frac{ \pi }{ 4 }
\end{equation}

Using~%
\eqref{eq:Tpar}
and
\eqref{eq:Tort}
one can obtain the Fourier-transform of the field on the external surface of the aperture (surface with coordinates $r^{ \prime } = \xi^{ \prime } \sin\alpha$ and $z^{ \prime } = - \xi^{ \prime } \cos\alpha$)
in the following form:
\begin{equation}
\begin{aligned}
H_{ \varphi^\prime }  \approx   T_{\parallel }   &
H_{\varphi^\prime }^{ i } ( r^\prime, \varphi^{ \prime },  z^{ \prime } )   , \\
E_{\xi^\prime} \approx H_{\varphi^\prime}  \cos\theta_t ,  
&  \  \  \  \ 
E_{\zeta^\prime} \approx H_{\varphi^\prime}  \sin\theta_t ;
\end{aligned}
\label{eq:parfields}
\end{equation}
\begin{equation}
\begin{aligned}
E_{ \varphi^\prime } \approx    T_{ \bot  }  &  
E_{ \varphi^\prime }^{ i } ( r^\prime, \varphi^\prime, z^\prime )  ,        \\
H_{ \xi^\prime}= - E_{\varphi^\prime}  \cos\theta_t,
&  \  \  \  \ 
E_{\zeta^\prime} \approx - E_{\varphi^\prime}  \sin\theta_t .
\end{aligned}
\label{eq:ortfields}
\end{equation}
\ 

%%%%%%%%%%%%%%%%%%%%%%%%
%%%%%%%%%%%%%%%%%%%%%%%%
\subsection{\label{subsec:2.2}  Aperture integrals}
%%%%%%%%%%%%%%%%%%%%%%%%
%%%%%%%%%%%%%%%%%%%%%%%%

Now we should write the general Stratton-Chu formulas~%
\cite{GVT19, TGV19, TGVG20_PRA, TGV21_JOSAB}
in the form which is convenient for further calculation in the case of considered target. 
We denote the cylindrical coordinates of the observation point $\vec R$ as $r, \varphi, z$, and the coordinates of the point on the aperture as $r', \varphi', z'$: 
$$
    r'=-z'\tan \alpha=\xi '\sin \alpha, \ \ \ \ 
    z'=-\xi '\cos \alpha.
$$
The distance from the observation point to the point on the aperture is 
\begin{multline}
\tilde{ R } { = } \left| \vec{  R} {-} \vec{ R }^{ \prime } \right|
{=} \\
{=}
\sqrt{{{r}^{2}} + \xi {{'}^{2}}\sin {{\alpha }^{2}} - 2r\xi '\sin \alpha \cos \tilde{\varphi}  + {{\left( \xi '\cos \alpha +  z \right)}^{2}} },
\label{eq:(Rtilde)}
\end{multline}
where $\tilde{\varphi} = \varphi '-\varphi$. 

Remind that Stratton-Chu formulas include the series of differentiation operations over the Green function $G(\tilde R)=\exp(i k \tilde R) / \tilde R$. To simplify the formulas, we introduce the following small restriction: we assume that the distance from the observation point to the aperture is much more than the wavelength under consideration, that is, for all points on the aperture $k \tilde R \gg 1$. Then, we can differentiate  the ``fast'' factor  $\exp(i k \tilde R)$ only, and consider  the ``slow''  factor  $1 / \tilde R$ as a constant. 

The corresponding transformations of aperture integrals \cite{GVT19,TGV19, TGVG20_PRA, TGV21_JOSAB} are cumbersome, but not very complicated. We give the final results for the electric field omitting all analytic calculations: 
\[\vec{E}\left( {\vec{R}} \right)={{\vec{E}}^{(h)}}\left( {\vec{R}} \right)+{{\vec{E}}^{(e)}}\left( {\vec{R}} \right),\]
\begin{widetext}
\begin{multline}\label{eq:exact1}
    E_{r}^{(h)}\left( {\vec{R}} \right)=\frac{ik}{4\pi }\int\limits_{\Sigma }{\left\{ -{{H}_{\varphi '}}\left[ rr'\sin \alpha {{\sin }^{2}}\tilde{\varphi }+\left( z'-z \right)\left( r'\cos \tilde{\varphi }-r \right)\cos \alpha +{{\left( z'-z \right)}^{2}}\sin \alpha \cos \tilde{\varphi } \right] \right.-}
    \\
    \left.-{{H}_{\xi '}}\left[ r{{'}^{2}}-rr'\cos \tilde{\varphi }+{{\left( z'-z \right)}^{2}} \right]\sin \tilde{\varphi } \right\}\frac{{{e}^{ik\tilde{R}}}}{{{{\tilde{R}}}^{3}}}d\Sigma '
\end{multline}
\begin{multline}\label{eq:exact2}
    E_{\varphi }^{(h)}\left( {\vec{R}} \right)\approx \frac{ik}{4\pi }\int\limits_{\Sigma }{\left\{ {{H}_{\xi '}}\left[ \left( {{r}^{2}}+r{{'}^{2}}+{{\left( z'-z \right)}^{2}} \right)\cos \tilde{\varphi }-2rr'\left( 1+{{\cos }^{2}}\tilde{\varphi } \right) \right] \right.}-
    \\
    \left. -{{H}_{\varphi '}}\left[ \left( {{r}^{2}}-rr'\cos \tilde{\varphi }+{{\left( z'-z \right)}^{2}} \right)\sin \alpha +r'\left( z'-z \right)\cos \alpha  \right]\sin \tilde{\varphi } \right\}\frac{{{e}^{ik\tilde{R}}}}{{{{\tilde{R}}}^{3}}}d\Sigma '
\end{multline}
\begin{equation}\label{eq:exact3}
E_{z}^{(h)}=\frac{ik}{4\pi }\int\limits_{\Sigma }{\left\{ {{H}_{\varphi '}}\left[ \left( {{r}^{2}}+r{{'}^{2}}-2rr'\cos \tilde{\varphi } \right)\cos \alpha +\left( z'-z \right)\left( r'-r\cos \tilde{\varphi } \right)\sin \alpha  \right]-{{H}_{\xi '}}r\left( z'-z \right)\sin \tilde{\varphi } \right\}\frac{{{e}^{ik\tilde{R}}}}{{{{\tilde{R}}}^{3}}}d\Sigma '}
\end{equation}
\begin{equation}\label{eq:exact4}
    \left( \begin{aligned}
  & E_{r}^{(e)} \\
 & E_{\varphi }^{(e)} \\
 & E_{z}^{(e)} \\
\end{aligned} \right)=\frac{ik}{4\pi }\int\limits_{\Sigma }{\left( \begin{aligned}
  & {{E}_{\xi '}}\left( z'-z \right)\cos \tilde{\varphi }-{{E}_{\varphi '}}\left[ r'\cos \alpha +\left( z'-z \right)\sin \alpha  \right]\sin \tilde{\varphi } \\
 & {{E}_{\xi '}}\left( z'-z \right)\sin \tilde{\varphi }+{{E}_{\varphi '}}\left[ \cos \alpha \left( r'\cos \tilde{\varphi }-r \right)+\left( z'-z \right)\sin \alpha \cos \tilde{\varphi } \right] \\
 & -{{E}_{\xi '}}\left( r'-r\cos \tilde{\varphi } \right)-{{E}_{\varphi '}}r\sin \alpha \sin \tilde{\varphi } \\
\end{aligned} \right)\frac{{{e}^{ik\tilde{R}}}}{{{{\tilde{R}}}^{2}}}d\Sigma ',}
\end{equation}

\end{widetext}
where
$\int\limits_{\Sigma }{d\Sigma '=}\sin \alpha \int\limits_{{{\xi }_{1}}}^{{{\xi }_{2}}}{d\xi '\int\limits_{0}^{2\pi }{d\varphi '\xi '}}$.
The integration limits
$ \xi_{ 1 } $,
$ \xi_{ 2 } $
are determined by the borders of the aperture: 
\begin{equation}
\xi_{ 1 } = a/\sin{\alpha}, \ \ \ 
\xi_{ 2 } = a/\sin{\alpha} + d.
\label{eq:(31)}
\end{equation}
%

%%%%%%%%%%%%%%%%%%%%%%%%%%%%%%%%%%%%%%%%%%%%%%%%%%%%%%%%%%%%%%%%%
%%%%%%%%%%%%%%%%%%%%%%%%%%%%%%%%%%%%%%%%%%%%%%%%%%%%%%%%%%%%%%%%%
\subsection{\label{subsec:2.3} Fraunhofer area }
%%%%%%%%%%%%%%%%%%%%%%%%%%%%%%%%%%%%%%%%%%%%%%%%%%%%%%%%%%%%%%%%%
%%%%%%%%%%%%%%%%%%%%%%%%%%%%%%%%%%%%%%%%%%%%%%%%%%%%%%%%%%%%%%%%%

The formulas \eqref{eq:exact1}--\eqref{eq:exact4} can be simplified in the Fraunhofer (or far-field) area where the wave parameter is large: 
$ D =  \frac{\lambda R}{\pi d^2} \gg 1$.
In this region the wave is a quasi-plane transverse one, exactly speaking, a spherical wave with small curvature  of the constant phase surface. Therefore it is convenient to use the spherical coordinates $R, \theta, \varphi$. The nonzero components of the field in Fraunhofer area are equal approximately to the following: 
\begin{widetext}
\begin{equation}
\label{eq:Frau1}
E_{\theta }^{{}}=\frac{ik{{e}^{ikR}}}{4\pi R}\int\limits_{\Sigma }{\left\{ -{{H}_{\varphi '}}\left[ \left( \cos {{\theta }_{t}}+\sin \alpha \cos \theta  \right)\cos \tilde{\varphi }+\cos \alpha \sin \theta  \right]+{{E}_{\varphi '}}\left( \sin \alpha +\cos {{\theta }_{t}}\cos \theta  \right)\sin \tilde{\varphi } \right\}{{e}^{ik\xi '\Phi }}d\Sigma '}, 
\end{equation}
\begin{equation}
\label{eq:Frau2}
E_{\varphi }^{{}}=\frac{ik{{e}^{ikR}}}{4\pi R}\int\limits_{\Sigma }{\left\{ -{{H}_{\varphi '}}\left( \sin \alpha +\cos {{\theta }_{t}}\cos \theta  \right)\sin \tilde{\varphi }-{{E}_{\varphi '}}\left[ \left( \cos {{\theta }_{t}}+\sin \alpha \cos \theta  \right)\cos \tilde{\varphi }+\cos \alpha \sin \theta  \right] \right\}}\,{{e}^{ik\xi '\Phi }}d\Sigma ',
\end{equation}
\end{widetext}
where
\[\Phi =\cos \alpha \cos \theta -\sin \alpha \sin \theta \cos \tilde{\varphi }.\] 
The wave in far-field area is a transverse with respect to the vector $\vec R$; correspondingly, $E_R$ component is absent. 

Note that, in the case $ r_0 = 0 $, the formulas \eqref{eq:exact1} -- \eqref{eq:exact4}, as well as \eqref {eq:Frau1} and  \eqref {eq:Frau2} are transformed into the corresponding formulas obtained in the paper~\cite {TGV19}.

%%%%%%%%%%%%%
\begin{figure*}
\includegraphics{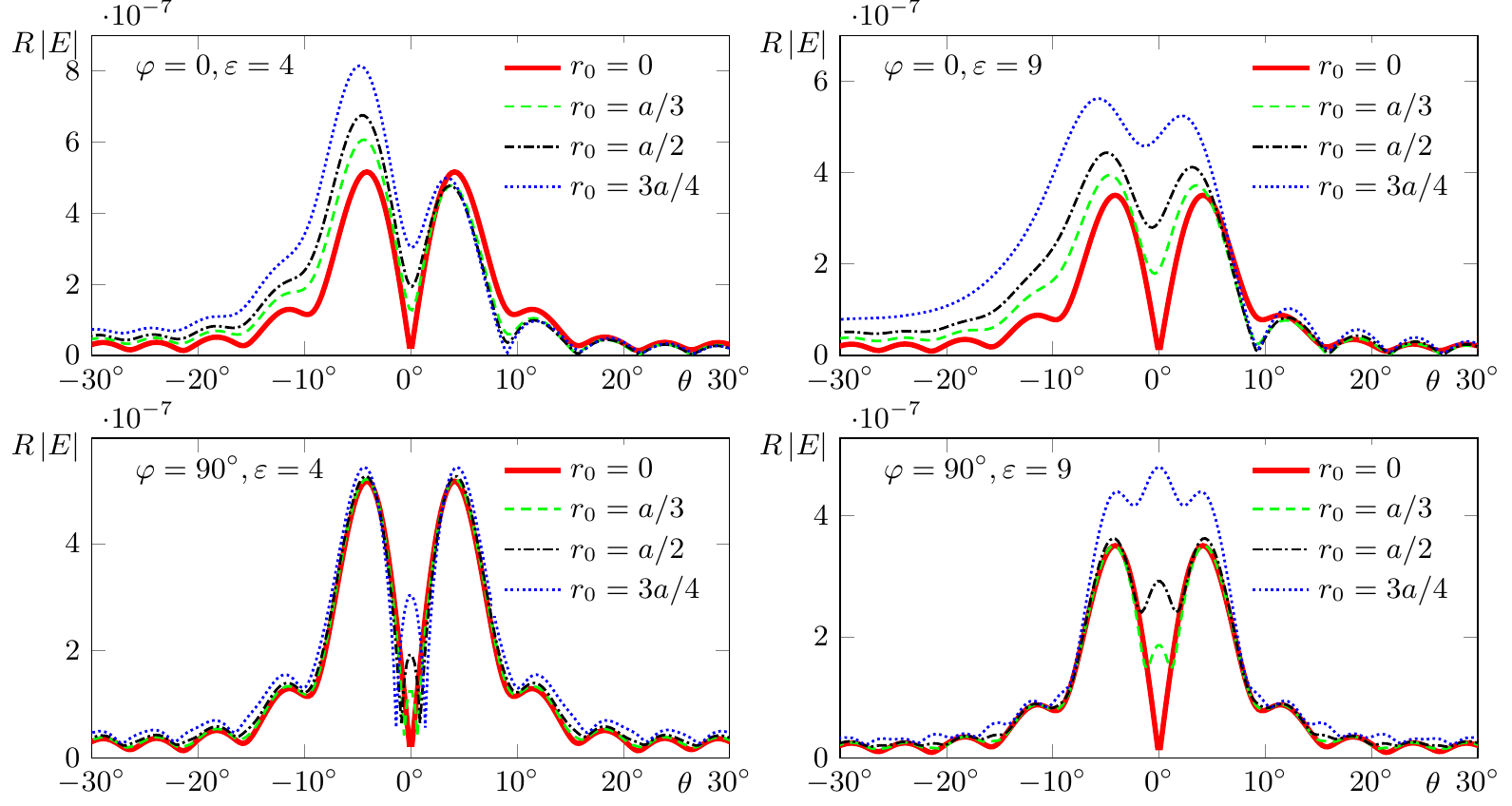}%
\caption{\label{fig:2}
The angular distribution of the magnitude of the electric field Fourier-transform in the Fraunhofer area for the velocity $\beta=\beta_*$ corresponding to the case of ``Cherenkov spotlight'' (in units $\mathrm{ V \cdot s }$). 
Parameters: $ a = c / \omega $,  $ d = 100 c / \omega = 100 / k $, $ q = 1~\mathrm{nC} $,
$ \mu=1 $,    $\alpha=20^{\rm o}$;  the permittivity $\varepsilon$, the charge shift $r_0$, and the observation plane angle $\varphi$
are indicated in the plots; $\beta=0.6726$ for $\varepsilon=4$ and $\beta=0.5384$ for $\varepsilon=9$. 
The negative values of $\theta$ correspond to value $\varphi$ equal $\pi$ plus the value $\varphi$ indicated on the plot.}
\end{figure*}
\begin{figure*}
\includegraphics{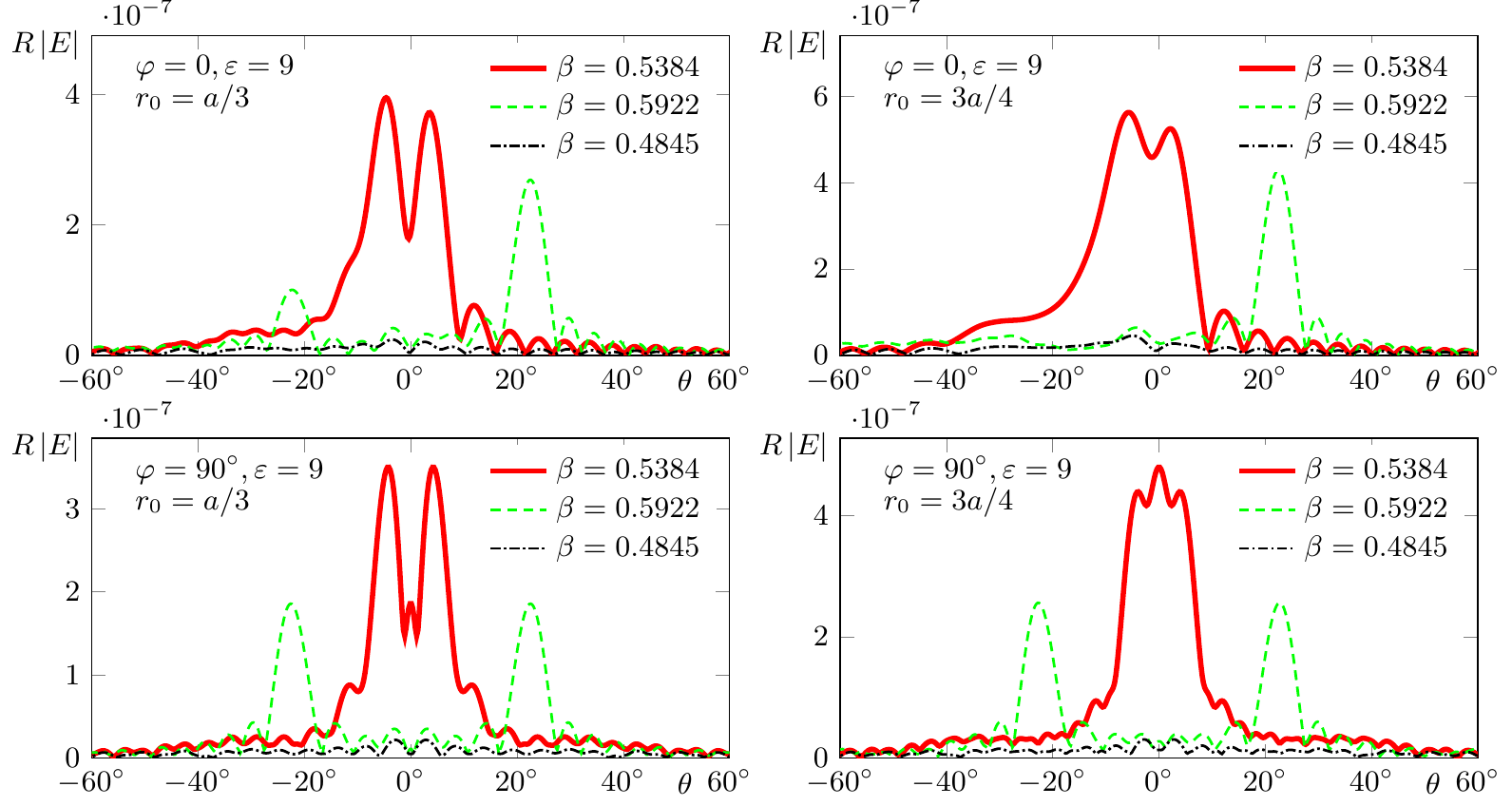}%
\caption{\label{fig:3}
The same as in Fig.~\ref{fig:2} for $\varepsilon=9$ and different velocities: the red lines correspond to the velocity $\beta_*$ (Cherenkov spotlight regime), the green dashed lines correspond to $\beta=1.1\beta_*$, the black dashed-dotted line correspond to $\beta=0.9\beta_*$.}
\end{figure*}
%%%%%%%%%%%%%%%

%%%%%%%%%%%%%%%%%%%%%%%%%%%%
%%%%%%%%%%%%%%%%%%%%%%%%%%%%
\subsection{\label{subsec:2.4} Results of computations}
%%%%%%%%%%%%%%%%%%%%%%%%%%%%
%%%%%%%%%%%%%%%%%%%%%%%%%%%%

Here we demonstrate the results of computation of the field in Fraunhofer area obtained with use of formulas \eqref{eq:Frau1}, \eqref{eq:Frau2}. First of all, it is interesting to analyze the influence of the charge off-axis motion on the ``Cherenkov spotlight'' effect which was described in~\cite{TGV19} for the case $r_0=0$. 

Figure~\ref{fig:2} shows the angle dependency of the field for different values of permittivity and  for different shifts of the charge trajectory $r_0$. It is assumed that the shift occurs in the side of positive values of $x$. 
The cross sections $\varphi=0, \pi$ and $\varphi=\pi/2, 3\pi/2$ are shown. 
It is assumed that the velocity $\beta$ corresponds to the case of ``Cherenkov spotlight'' effect for $r_0=0$ (this velocity is indicated as $\beta_*$). 
The vertical axis on the plots shows the value 
$ R |E| $
which does not depend on the distance
$ R $
in the Fraunhofer area.
For convenience the negative values of $\theta$ in the figure correspond to positive ones for $\varphi$ shifted on $\pi$.

Each plot contains four curves. For the bold red solid curve, the shift of the charge is zero
(i.e. $ r_0 = 0 $); it illustrates the ``Cherenkov spotlight'' effect in symmetrical case.
Other curves correspond to the cases when $ r_0 \ne 0 $. One can see that the spotlight effect is retained even when there is a large shift. Moreover, in the plane $\varphi=0, \pi$, the radiation in the case $r_0 \ne 0$ is usually even more intense than in the case of $r_0 = 0$. Most significant increase of radiation  occurs in the direction opposite to the shift of the charge trajectory (that is, in the region of the negative angles $\varphi$). These effects can be explained by the fact that the excitation of Cherenkov radiation increases as the  trajectory approaches the channel boundary. Another effect is that the field is not equal to zero on the structure axis in the case of $r_0 \ne 0$. Moreover, the field has a maximum on the symmetry axis in the cross section $\varphi=\pi/2, 3 \pi/2$.

Besides the role of the trajectory offset, it is interesting to evaluate the effect of the velocity deviation from the "spotlight" velocity $\beta_*$. This effect is illustrated in Fig.~\ref{fig:3}, which shows the radiation amplitude at the velocity deviation of 10\% from $\beta_*$. As you can see, this influence is very great. For $\beta=1.1\beta_*$, the maximums shift and decrease essentially (green dashed lines), and for $\beta=0.9\beta_*$, the spotlight effect practically disappears (black dash-dotted lines).

%%%%%%%%%%%%%%%
\begin{figure}
\centering
\includegraphics{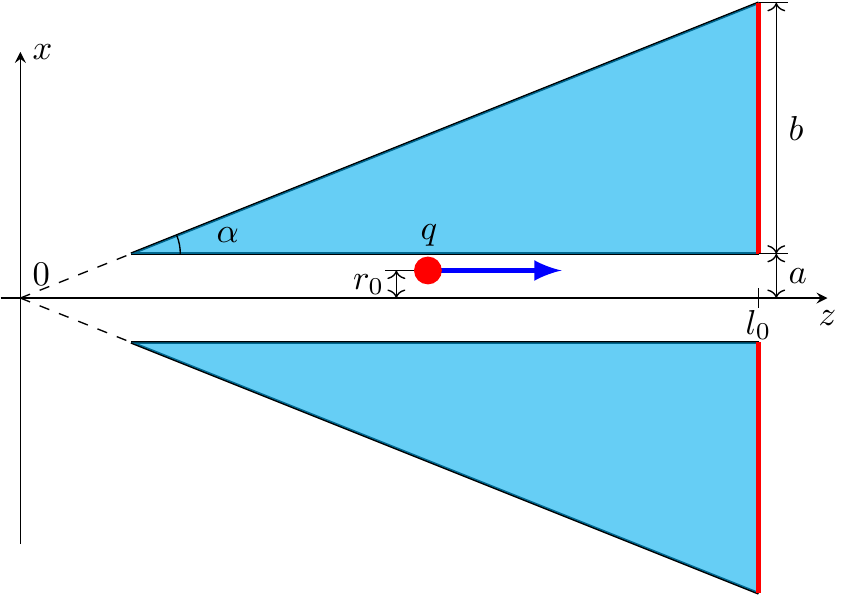}
\caption{\label{fig:4}%
The ``reversed'' cone cross-section.
}
\end{figure}
%%%%%%%%%%%%%%%
%%%%%%%%%%%%%%%%%%%%%%%%%%%%%%%%%%%%%%%%%
\begin{figure} 
\centering
%[width=0.85\linewidth]
\includegraphics{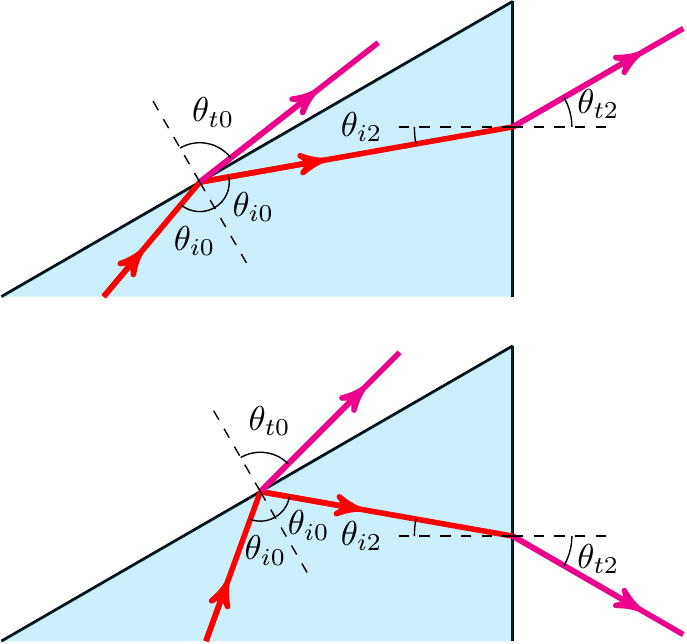}
\caption{\label{fig:5}%
The path of the ray for the case
${{\theta }_{i}}>0$
(top) and
${{\theta }_{i}}<0$
(bottom).
}
\end{figure}
%%%%%%%%%%%%%%%%%%%%%%%

%%%%%%%%%%%%%%%%%%%%%%%%%%%%%%%%%
%%%%%%%%%%%%%%%%%%%%%%%%%%%%%%%%%
\section {\label{sec:3} The case of ``inverted'' cone }
%%%%%%%%%%%%%%%%%%%%%%%%%
%%%%%%%%%%%%%%%%%%%%%%%%%
\subsection{\label{subsec:3.1}  The field on the aperture}
%%%%%%%%%%%%
%%%%%%%%%%%%%

Let us now consider the problem with charge movement from the top of the cone to its base (Fig.~\ref{fig:4}). 
The radius of the cone base is $ b + a $, and the cone angle is 
$ \alpha $. Accordingly, the length of the target along its axis is
$ l = b \cot \alpha $ and the distance from the top of the cone to its base is
$ l_0 = ( a + b ) \cot \alpha $.
As earlier, we assume here that the target size is much larger than the  wavelength under consideration. 
Note that the axially symmetric case ($r_0=0$) has been analyzed by us in the paper \cite{TGVG20_PRA}. 

The ``etalon'' problem for this situation is the same as in the Section 2 (this is the problem with the channel in the unbounded medium). 
Therefore the ``initial'' CR wave in the bulk of the target is determined by the formulas \eqref{eq:Hphiappr}, \eqref{eq:Ephiappr}. 

We are mainly interested in the Cherenkov spotlight regime. 
As we have shown for the case when $ r_0 = 0 $~\cite{}, this effect can be realized with use of the wave which is once reflected from the cone generatrix~\cite{TGVG20_PRA} and then exits the target from its base surface. Therefore, the cone base serves as the aperture in this case (solid red line in Fig.~\ref{fig:4}). 
We recall here the way of this wave (Fig.~\ref{fig:5}). This wave falls on the cone generatrix at the angle 
\[
\theta_{ i 0 } = \pi / 2 + \alpha - \theta_p 
\ \ \ 
(  {\rm where} \ \theta_p = \arccos \left[ 1 / \left( n \beta \right) \right]),
\]
and reflects from it at the same angle $\theta_{ r 0 }=\theta_{ i 0 }$. Then this wave incidents the cone base at the angle 
\begin{equation}
\theta_{ i 2 }
=
\theta_{ r 0 } - \left( \pi /  2 - \alpha \right)
= 2 \alpha - \theta_p,
\label{eq:thetai2}
\end{equation}
and refracts at the angle  
\begin{equation}
\theta_{ t 2 }
=\mathrm{ arcsin }( n \sin \theta_{ i 2 } ).
\label{eq:(21)}
\end{equation}

Cherenkov spotlight effect occurs if $\theta_{ t 2 }=0$. 
Thus, the wave experiences one reflection and one refraction in that case.
For the wave with ``parallel'' polarization ($ \parallel $) containing the components
$ E_{ z \omega } $, $ E_{ r \omega } $ and $ H_{ \varphi \omega } $, 
the reflection coefficient (for the reflection at the cone generatrix) is
\begin{equation}
R_{ \parallel 0 } =
\frac{ \cos \theta_{ i 0 } - \sqrt{ \varepsilon } \cos \theta_{ t 0 } }
{ \cos \theta_{ i 0 } + \sqrt{ \varepsilon } \cos \theta_{ t 0 } },
\label{eq:Rpar2}
\end{equation}
and the transmission coefficient (for the transmission at the cone base surface) is
\begin{equation}
\label{eq:Tpar2}
T_{ \parallel 2} = 
\frac{ 2 \cos \theta_{ i 2 } }{ \cos \theta_{ i 2 } + \sqrt{ \varepsilon } \cos \theta_{ t 2 } }.
\end{equation}
For the wave with ``orthogonal'' polarization ($ \bot $) containing the components
$ H_{ z \omega } $, $ H_{ \rho \omega } $ and $ E_{ \varphi \omega } $,
corresponding coefficients are

\begin{equation}
R_{ \bot 0 } =
\frac{ \sqrt{ \varepsilon } \cos \theta_{ i 0 } - \cos \theta_{ t 0 } }
{ \sqrt{ \varepsilon } \cos \theta_{ i 0 } + \cos \theta_{ t 0 } }.
\label{eq:Rort2}
\end{equation}
\begin{equation}
\label{eq:Tort2}
T_{ \bot 2 } =
\frac{ 2 \sqrt{ \varepsilon } \cos \theta_{ i 1 } }{ \sqrt{ \varepsilon } \cos \theta_{ i 1 } + \cos \theta_{ t 1 } }.
\end{equation}
%

%%%%%%%%%%%%%%%%%%%%%%%%%%%%%%%%%%%%%%%%%%%%%%%%%%%%%%%%%%%%%%%%%%%%%%%%%%%%%%%%%%%%%%%%%%%%

Omitting  intermediate transformations, we now present the result for the Fourier transforms of the field components at the aperture at the point with cylindrical coordinates $ r^{ \prime } $, $ \varphi^{ \prime } $, $ z^{ \prime } = l_0 + 0 $:
\begin{equation}
\begin{aligned}
\left. H_{ \varphi^{ \prime } } \right|_{ z^{ \prime } = l_0 + 0 }
&\approx
R_{ \parallel 0 } T_{ \parallel 2 } H_{ \varphi^{\prime}}^{ i } ( r^{ \prime }, \varphi^{ \prime },  l_0 )
, \\
E_{ r^{ \prime } }
&\approx
H_{ \varphi^{ \prime } }
\cos \theta_{ t 2 }, \\
E_{ z^{ \prime } }
&\approx
- H_{ \varphi^{ \prime } }
\sin \theta_{ t 2 };
\end{aligned}
\label{eq:parfields-rev}
\end{equation}
\begin{equation}
\begin{aligned}
\left. E_{ \varphi^{ \prime } }\right|_{ z^{ \prime } = l_0 + 0 }
&\approx
R_{ \bot 0 } T_{ \bot 2 }  E_{ \varphi^{\prime}}^{i} ( r^{ \prime }, \varphi^{ \prime },  l_0 ) , \\
H_{ r^{ \prime } }
&\approx
-E_{ \varphi^{ \prime } }
\cos \theta_{ t 2 }, \\
H_{ z^{ \prime } }
&\approx
E_{ \varphi^{ \prime } }
\sin \theta_{ t 2 }.
\end{aligned}
\label{eq:ortfields-rev}
\end{equation}
The expressions for $H_{ \varphi^{\prime}}^{i} $ and $E_{ \varphi^{\prime}}^{i} $ are formally coincide with \eqref{eq:Hphiappr}, \eqref{eq:Ephiappr} however the formula for $\Phi_i$ is different. Calculation give the following result: 
\begin{equation}
\begin{aligned}
& \Phi_ i ( r^\prime, z^\prime) =
k n \left( r^\prime \sin \theta_{ i 2 } + z^\prime \cos \theta_{ i 2 } \right) - \pi / 4
= \\
&  =  k r^\prime \sin \theta_{ t 2 } + k n z^\prime \cos \theta_{ i 2 } - \pi / 4,
\end{aligned}
\label{eq:Phii2)}
\end{equation}

%
%%%%%%%%%%%%%%%%%%%%%%%
%%%%%%%%%%%%%%%%%%%%%%%
\subsection{\label{subsec:3.2}  Aperture integrals}
%%%%%%%%%%%%%%%%%%%%%%%
%%%%%%%%%%%%%%%%%%%%%%%
Let us write the general Stratton-Chu formulas~%
\cite{GVT19,TGV19, TGVG20_PRA, TGV21_JOSAB}
in the form which is convenient for further calculation. 
Remind that we denote the cylindrical coordinates of the observation point $\vec R$ as $r, \varphi, z$, and the coordinates of the point on the aperture $\vec R'$ as $r', \varphi', z'$.  
The distance from the observation point to the point on the aperture is 
$
\tilde R = \sqrt{ r^2 + { r'}^2 - 2 r r' \cos\tilde\varphi + \tilde z^2 },
$
where $ \tilde \varphi = \varphi' - \varphi $,   $ \tilde z = z - l_0 $. 

As in the previous section, we assume, that the distance from the observation point to any point on the aperture is much more than the wavelength under consideration: $k \tilde R \gg 1$. It allows to differentiate only the ``fast'' factor  $\exp(i k \tilde R)$ in the Green function considering the ``slow''  factor  $1 / \tilde R$ as a constant. 

We give the final result for the electric field Fourier transform omitting all transformations: 
\[\vec{E} ( {\vec{R}} ) = {{\vec{E}}^{(h)}} ( {\vec{R}} ) + {{\vec{E}}^{(e)}} ( {\vec{R}} ),\]

\begin{widetext}
\begin{equation} \label{Erev1}
E_{r}^{(h)} = - \frac{ik}{4\pi }
\int\limits_{\Sigma } \left[
   H_{r'} \sin\tilde\varphi \left( {r'}^2 - r r'  \cos \tilde\varphi + {{\tilde z}^{2}} \right) 
  + {{H}_{\varphi'}}\left( rr'{{\sin }^{2}}\tilde\varphi + {{\tilde z}^{2}}\cos \tilde\varphi \right)
  \right]
\frac{{{e}^{ik\tilde{R}}}}{{{{\tilde{R}}}^{3}}}d\Sigma ' ,
\end{equation}
\begin{equation}\label{Erev2}
  E_{r}^{(h)} = \frac{ik}{4\pi }\int\limits_{\Sigma }     \left\{ 
  {{H}_{r'}}\left[ \left( {{r}^{2}}+r{{'}^{2}}+{{\tilde z}^{2}} \right) \cos \tilde\varphi 
                 - r r' \left( 1+{{\cos }^{2}} \tilde\varphi \right] \right)
   - {{H}_{\varphi '}}\sin \tilde\varphi \left[ {{r}^{2}} - r r' \cos \tilde\varphi + {{\tilde z}^{2}} \right]
  \right\}      \frac{{{e}^{ik\tilde{R}}}}{{{{\tilde{R}}}^{3}}}d\Sigma ',
\end{equation}
\begin{equation}\label{Erev3}
  E_{r}^{(h)} = - \frac{i k \tilde z }{4\pi }    \int\limits_{\Sigma } 
  \left[ {{H}_{\varphi '}}\left( r'-r\cos \tilde\varphi \right)-{{H}_{r'}}r\sin \tilde\varphi \right]
  \frac{{{e}^{ik\tilde{R}}}}{{{{\tilde{R}}}^{3}}}d\Sigma ',
\end{equation}
\begin{equation}\label{Erev3}
\left\{ \begin{aligned}
  & E_{r}^{(e)} \\
 & E_{\varphi }^{(e)} \\
 & E_{z}^{(e)} \\
\end{aligned} \right\}  =  - \frac{ik}{4\pi }\int\limits_{\Sigma }
{\left\{ \begin{aligned}
  &  \left( {{E}_{r'}}\cos \tilde\varphi  -  {{E}_{\varphi '}}\sin \tilde\varphi \right) \tilde z \\
 &  \left( {{E}_{r'}}\sin \tilde\varphi + {{E}_{\varphi '}}\cos \tilde\varphi \right)  \tilde z \\
 & {{E}_{r'}}\left( r'-r\cos \tilde\varphi \right) + {{E}_{\varphi '}}r\sin \tilde\varphi \\
\end{aligned} \right\}
\frac{{{e}^{ik\tilde{R}}}}{{{{\tilde{R}}}^{2}}}d\Sigma ',}
\end{equation}
where 
$ \int\limits_{\Sigma } d\Sigma ' = \int\limits_{a}^{a+b}  r'dr' \int\limits_{0}^{2\pi}d\varphi' $.

%%%%%%%%%%%%%%%%%%%%%%%
%%%%%%%%%%%%%%%%%%%%%%%
\subsection{\label{subsec:3.3} Fraunhofer area }
%%%%%%%%%%%%%%%%%%%%%%%
%%%%%%%%%%%%%%%%%%%%%%%

The formulas \eqref{Erev1}--\eqref{Erev3} can be simplified in the Fraunhofer (or far-field) area where the wave parameter is large: $ D =  \lambda R / (\pi d^2) \gg 1$. 
In this region, the wave is a spherical transverse wave with small curvature of constant phase surface. Using the spherical coordinates $R, \theta, \varphi$ one can write the nonzero components of the field in Fraunhofer area in the following form:  
\begin{equation}\label{FRRev_theta}
  E_\theta = \frac{ i k \exp \left( ikR - i k l_0 \cos \theta \right)} {4\pi R} 
  \int\limits_{\Sigma }   \left[ E_{\varphi '} \left( 1+ \cos \theta_t \cos \theta  \right)\sin \tilde\varphi 
    - H_{\varphi '} \left( \cos \theta_t + \cos \theta  \right) \cos \tilde\varphi \right] 
  e^{ - i k r' \sin \theta \cos \tilde\varphi } d\Sigma',
\end{equation}
\begin{equation}\label{FRRev_varphi}
E_\varphi = \frac{ i k \exp \left( i k R - i k l_0 \cos \theta  \right)} {4\pi R} 
\int\limits_{\Sigma }\left[ - E_{\varphi'} \left( \cos\theta_t + \cos \theta  \right) \cos\tilde\varphi 
- H_\varphi' \left( 1 + \cos\theta_t \cos\theta  \right) \sin\tilde\varphi \right]
e^{ - i k r' \sin \theta \cos \tilde\varphi } d\Sigma'.
\end{equation}
\end{widetext}
If $ r_0 = 0 $ then the formulas \eqref {Erev1} -- \eqref {Erev3}, as well as \eqref {FRRev_theta} and \eqref {FRRev_varphi} are transformed into the corresponding formulas obtained in the paper~\cite {TGVG20_PRA}.

%%%%%%%%%%%%%%%%%%%%%%%%%
\begin{figure*}
\includegraphics{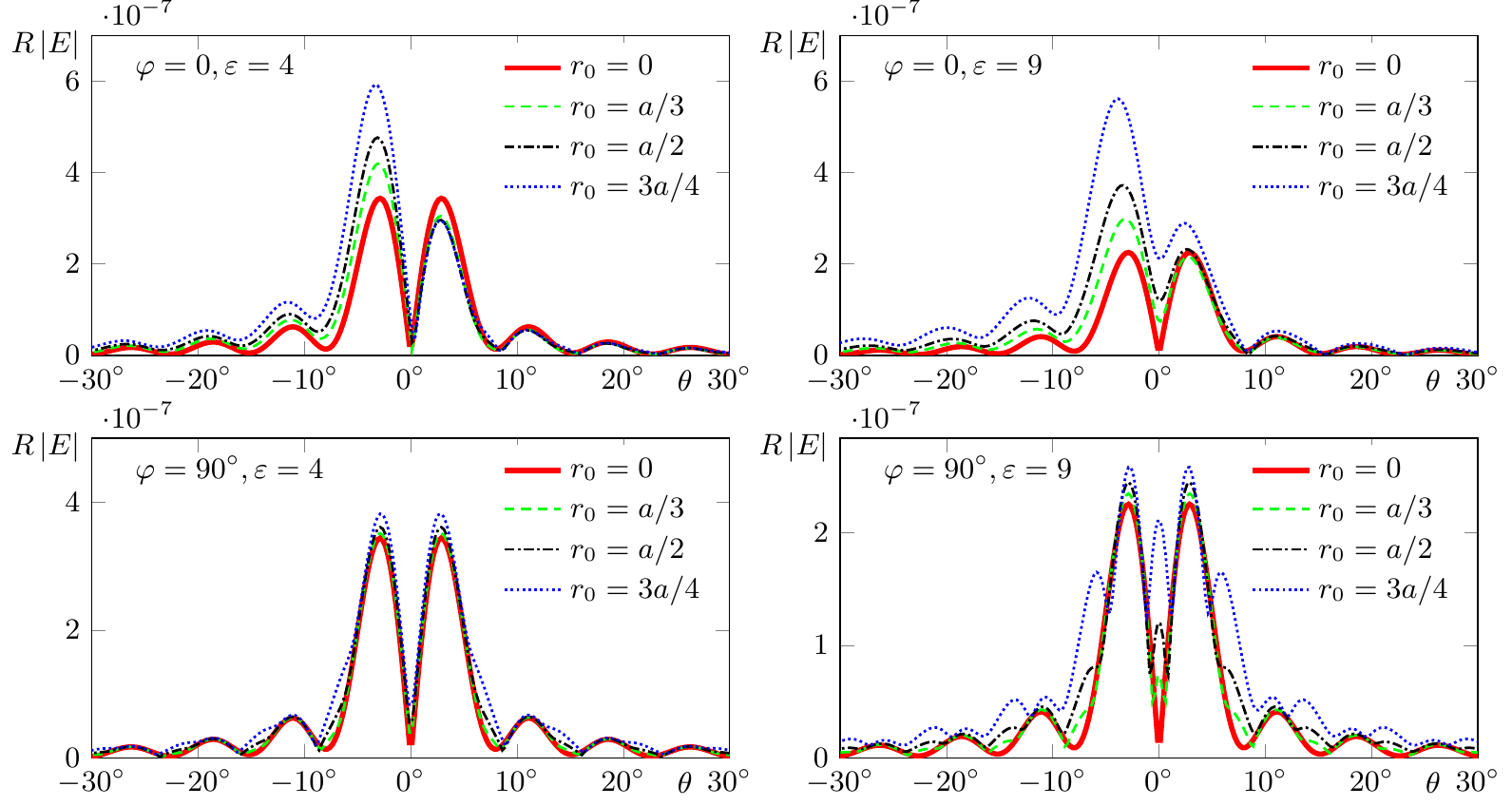}%
\caption{\label{fig:6}
The angular distribution of the magnitude of the electric field Fourier-transform in the Fraunhofer area for the velocity $\beta=\beta_*$ corresponding to the case of ``Cherenkov spotlight'' (in units $\mathrm{ V \cdot s }$). 
Parameters: $ a = c / \omega $,
$ b = 50 c / \omega = 50 / k $,
$ q = 1~\mathrm{nC} $,
$ \mu=1 $,   $\alpha=20^{\rm o }$;  
the charge shift $r_0$, the permittivity $ \varepsilon $, and the observation angle $\varphi$ are indicated in the plots; $\beta=0.6527$ for $\varepsilon=4$ and $\beta=0.4351$ for $\varepsilon=9$. 
The negative values of $\theta$ correspond to value $\varphi$ equal $\pi$ plus the value $\varphi$ indicated on the plot.}
\end{figure*}
\begin{figure*}
\includegraphics{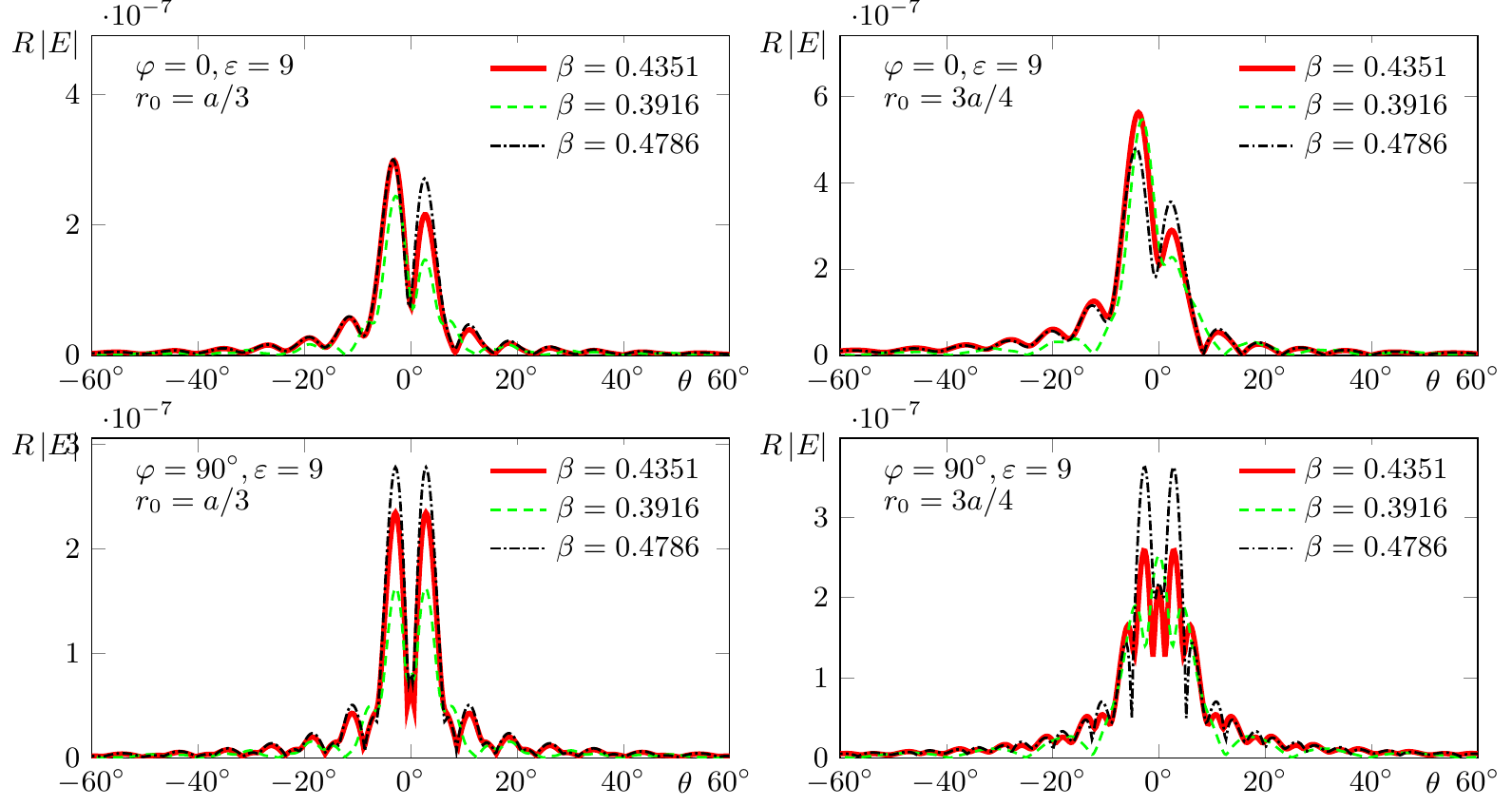}%
\caption{\label{fig:7}
The same as in Fig.~\ref{fig:6} for $\varepsilon=9$ and different velocities: the red lines correspond to the velocity $\beta_*$ (Cherenkov spotlight regime), the green dashed lines correspond to $\beta=1.1\beta_*$, the black dashed-dotted line correspond to $\beta=0.9\beta_*$.}
\end{figure*}
%%%%%%%%%%%%%%%%%%%%%%%%%

%%%%%%%%%%%%%%%%%%%%%%%%
%%%%%%%%%%%%%%%%%%%%%%%
\subsection{\label{subsec:3.4} Results of computations }
%%%%%%%%%%%%%%%%%%%%%%%%%
%%%%%%%%%%%%%%%%%%%%%%%%

The results of computation of the field obtaining by the formulas \eqref{FRRev_theta}, \eqref{FRRev_varphi} are shown in Fig.~\ref{fig:6}, \ref{fig:7}. First of all, we demonstrate the influence of the charge off-axis motion on the ``Cherenkov spotlight'' effect which was described in~\cite{TGVG20_PRA} for the case $r_0=0$. 

Figure~\ref{fig:6} shows the angular dependency of the field for different values of permittivity and  for different shifts of the charge trajectory $r_0$. It is assumed that the shift occurs in the side of positive values of $x$. 
The cross sections $\varphi=0, \pi$ and $\varphi=\pi/2, 3\pi/2$ are shown. It is assumed that the velocity $\beta$ corresponds to the case of ``Cherenkov spotlight'' effect for $r_0=0$ (this velocity is indicated as $\beta_*$). 
The vertical axis on the plots shows the value of $ R |E| $
which does not depend on the distance $ R $ in the Fraunhofer area. 
For convenience the negative values of $\theta$ in the figure correspond to positive ones for $\varphi$ shifted on $\pi$. For the bold red solid curves, the shift of the charge is zero
($ r_0 = 0 $); it illustrates the ``Cherenkov spotlight'' effect in symmetrical case.
Other curves correspond to the cases when $ r_0 \ne 0 $.

The main conclusions here are the same as in the case of the straight cone, namely:

- The spotlight effect is retained even when there is the large shift of the charge trajectory; moreover, in the plane $\varphi=0, \pi$, the radiation in the case $r_0 \ne 0$ is usually even more intense than in the case of $r_0 = 0$;

- Most significant increase of radiation occurs in the direction opposite to the shift of the charge trajectory; 

- The field is not equal to zero on the structure axis in the case of $r_0 \ne 0$. 

The influence of the velocity deviation on the spotlight regime is illustrated in Fig.~\ref{fig:7} which shows the field value at the velocity differing in 10\% from the ``spotlight velocity'' $\beta_*$. It is interesting that this influence is much less than in the case of the straight cone (compare Fig.~\ref{fig:7} and Fig.~\ref{fig:3}). The spotlight effect is fully retained when the speed is changed by 10\%. Moreover, under certain conditions, the radiation is obtained even more than with $\beta=\beta_*$. 

Thus, the inverted cone offers significant advantages for realizing the spotlight effect over the straight cone. These schemes are equally insensitive to  variations in the displacement of the trajectory from the axis. However the radiation from the inverted cone is much more stable with respect to variations in the charge velocity. Earlier, one more important advantage of the inverted cone  was also noted ~\cite{TGVG20_PRA}: with such a scheme, it is always possible to select the cone parameters, which would make it possible to realize the spotlight effect  (including the case of  ultrarelativistic charge). For the straight cone, there are significant limits on the charge speed at which the spotlight effect is achievable.

%%%%%%%%%%%%%%%%%%%%%%%%%%%%%%%%%%%%%%%%%%%%%%%%%%%%%%%%%%%%%%%%%
\section{Conclusion}
%%%%%%%%%%%%%%%%%%%%%%%%%%%%%%%%%%%%%%%%%%%%%%%%%%%%%%%%%%%%%%%%%

We investigated Cherenkov radiation generated by charge moving in channel in the dielectric cone. Unlike  previous works, a shift of the charge trajectory from the symmetry axis was taken into account. We applied the aperture method which was developed by us earlier. The most attention was given to phenomenon of ``Cherenkov spotlight'' which was reported earlier for axially symmetrical problems. 
This effect allows essential enhancement of Cherenkov radiation intensity in the far-field region by proper selection of the target's parameters. 

We analyzed the influence of the shift of the charge trajectory from the symmetry axis on the spotlight regime, as well as the effect of the charge velocity variation. The main conclusions consist in the following. The spotlight effect is retained even for large displacement of trajectory from the symmetry axis both in case of the straight cone and in case of the inverted cone. The radiation can be even more than for the central  motion of charge. The most significant increase of radiation occurs in the direction opposite to the shift of the charge trajectory. However the radiation from the inverted cone is much more stable with respect to variations in the charge velocity. It is important as well that the inverted cone allows reaching the spotlight phenomenon for any speed of charge (due to selection of the cone parameters), while, for the straight cone, there are significant limits on the charge velocity which allow reaching the spotlight regime. 

%%%%%%%%%%%%%%%%%%%%%%%%%%%%%%%%%%%%%%%%%%%%%%%%%%%%%%%%%%%%%%%%%
%%%%%%%%%%%%%%%%%%%%%%%%%%%%%%%%%%%%%%%%%%%%%%%%%%%%%%%%%%%%%%%%%
\section{Acknowledgements}
%%%%%%%%%%%%%%%%%%%%%%%%%%%%%%%%%%%%%%%%%%%%%%%%%%%%%%%%%%%%%%%%%
%%%%%%%%%%%%%%%%%%%%%%%%%%%%%%%%%%%%%%%%%%%%%%%%%%%%%%%%%%%%%%%%%

This research was supported by the Russian Science Foundation, Grant No.~18-72-10137.

%\bibliography{SNG_Bibliography_Jun2021}
%

\end{document}